\documentclass[amsmath,amssymb,article,onecolumn]{revtex4}
\usepackage[english]{babel}
\usepackage{multirow}
\usepackage{graphicx}
\usepackage{dcolumn}
\usepackage{bm}

\newcommand{\be}{\begin{equation}}
\newcommand{\ee}{\end{equation}}
\newcommand{\bea}{\begin{eqnarray}}
\newcommand{\eea}{\end{eqnarray}}
\newcommand{\ba}{\begin{eqnarray*}}
\newcommand{\ea}{\end{eqnarray*}}

\newcommand{\Tr}{\mathrm{Tr}}

\newcommand{\lef}{\left(}
\newcommand{\rr}{\right)}

\begin{document}

\title[Beyond the Born-Oppenheimer approximation]
{Finite temperature electronic simulations beyond the Born-Oppenheimer approximation}

\author{Guglielmo Mazzola}
\email{gmazzola@sissa.it}
 \affiliation{International School for Advanced Studies (SISSA), and CRS Democritos, CNR-INFM, - Via Bonomea 265, I-34136 Trieste, Italy}

\author{Andrea Zen}
\email{andrea.zen@uniroma1.it}
\affiliation{Dipartimento di Fisica, Universit\`a di Roma ``La Sapienza'', Piazzale Aldo Moro 2, I-00185 Rome, Italy}

\author{Sandro Sorella}
\email{sorella@sissa.it}
\affiliation{International School for Advanced Studies (SISSA), and CRS Democritos, CNR-INFM, - Via Bonomea 265, I-34136 Trieste, Italy}

             
\begin{abstract}
We introduce a general technique  to compute finite temperature electronic properties by  a novel covariant formulation of the electronic partition function.
By using a rigorous variational upper bound to the  free energy 
we are led to the evaluation of a  
partition function that can be computed stochastically  by sampling  
electronic wave functions and atomic positions (assumed classical). 
In order to achieve this target we show that it is extremely important to consider the non trivial geometry of the space 
defined by the wave function ansatz.   
The method can be extended to any technique capable to provide an energy 
value over a given wave function ansatz depending on several 
variational parameters and atomic positions. In particular we can 
take into account electronic correlation, by using the standard variational quantum Monte Carlo method, that has been so far limited to zero temperature ground state properties.  
We show that our approximation 
reduces correctly to the standard Born-Oppenheimer (BO) one at zero  
temperature and to the correct high temperature limit.
At large enough temperatures this method allows to improve 
the BO, providing lower values of the 
electronic free energy, because within this method it is possible to 
take into account  the electron entropy.
We test this new method on the simple hydrogen molecule, where at low 
temperature we recover the correct   
BO  low temperature  limit.
Moreover, we show that 
the dissociation of the molecule is possible at a temperature 
much smaller than the BO prediction.
Several extension of the proposed technique  are also discussed, as for instance the calculation of critical (magnetic, superconducting) temperatures, 
or transition rates  in chemical reactions.

\end{abstract}

    \maketitle
\section{Introduction}
 The calculation of finite temperature electronic properties is one of the most important and challenging aspects of the numerical simulations.
In the past several progress have been done  by extending the 
DFT method to finite temperature \cite{GRUMBACH:1994p22136,PhysRev.137.A1441} or by using quantum Monte Carlo \cite{Foulkes:2001p19717}  (QMC) within various  path integral formulations \cite{Ceperley:1995p19543,Ceperley:1996p28593,Carlson:1999tl,Zhang:1999vc,Boninsegni:2006p24344,Pierleoni:2006p28485}, 
especially in the study of the hydrogen phase diagram \cite{Liberatore:2011dt,Liberatore:2011p28603,Morales:2010p28600,Delaney:2006p28594,Holzmann:2005ce,Pierleoni:2004p28607,Magro:1996p28605,Pierleoni:1994wc}.
In both cases many problems remain as for instance the lack of an accurate 
local functional at finite temperature for DFT methods prevents  
so far practical applications, and, within QMC techniques, 
the difficulty to deal with the fermion sign problem\cite{Ceperley:1991p19568}, restricts the spectrum of applicability to very few cases and very limited temperature ranges. 
On the other hand it is clear that, in many physical phenomena, such as the 
occurrence of magnetic or insulating  phases below a critical temperature, 
the electronic entropy cannot be neglected, even when the small ratio $\lambda_{ei}$  between 
the electronic mass and the atomic one, allows the decoupling of 
the electronic degrees of freedom from the atomic ones, within an acceptable approximation. 
In this paper we aim to extend the validity of the Born-Oppenheimer 
approximation in the following sense.
By using the smallness of $\lambda_{ei}$ we are generally lead 
to compute an electronic  partition function $Z[\bf R]$ at fixed nuclei position:
\begin{eqnarray}
Z&=& \int  d {\bf R}  ~ Z[{\bf R}] \label{firstbo} \\
Z[{\bf R}] &=& \Tr \exp ( - H_{\bf R}/T )  
\end{eqnarray} 
where $T$ is the temperature
 (here and henceforth the Boltzman constant is assumed to be one and we 
neglet for simplicity the overall constant coming from integration of the 
atomic momenta), $H_{\bf R}$ is the standard electronic Hamiltonian, 
that includes also the classical ionic contribution, and that depends only 
parametrically upon  the atomic positions ${\bf R}$. 
Eq.(\ref{firstbo})  is the first step of the Born-Oppenheimer approximation 
that- we remark- is generally valid for $\lambda_{ei}$ small, namely 
when the temperature $T$ is large enough that quantum effects on protons 
can be neglected. The 
second approximation, usually adopted within the BO approximation,  
 is to assume that the electronic degrees of freedom have a gap 
much larger than the temperature $T$ so that $Z[{\bf R}]$ can be approximated 
by $\exp ( - E_0 ({\bf R})/T) $ where $E_0 (\bf R) $ is the ground state energy 
of the hamiltonian $H_{\bf R}.$
In the following derivation we want to avoid the latter approximation, 
because, as emphasized before, in several cases 
it  may fail even when we are in the limit of small 
$\lambda_{ei}$. 
For instance the occurrence of a broken symmetry phase 
often  implies gapless electronic excitations 
in  $H_{\bf R}$, and the approximation 
$Z[{\bf R}]= \exp (  - E_0 ({\bf R})/T) $ cannot  be safely assumed. 
Other examples are conical intersections\cite{worth,Longuet,mead}, when for some particular 
ionic positions $H_{\bf R}$ becomes gapless and nearby the proximity 
between 
different (namely corresponding to low-lying excited states) BO energy surfaces 
is possible. In this conditions a pure 
electronic ground state technique
fails as the tunneling between different BO energy surfaces cannot be taken 
into account consistently.
As the last very important example
we mention the calculation of transition rates in chemical reactions, 
that cannot be accurately computed within a pure BO approximation\cite{butler,kash3463,Kuppermann1993577,MarkWu1991319,RevModPhys66}.

The main task of this paper is to device a method, able to quantify 
finite temperature properties of realistic systems, within a rigorous 
variational upper bound of the total free energy $F= -T \ln Z $, in the 
limit of small $\lambda_{ei}$.
The method we propose is supposed to be simple enough to avoid most 
of the known drawbacks, as does not rely on the knowledge of any particular 
functional, or, within our variational 
approximations, can be employed by quantum Monte 
Carlo, without facing the so called ``fermion sign problem''.

The paper is organized as follows.
The derivation of the approximate expression of the electronic partition function introduced and used in this work is given in section~\ref{s_upperbound}, and some important but more detailed aspects are reported in appendixes~\ref{app1}, \ref{app2}, \ref{app_comparisonBO} and \ref{app4}. 
This derivation is not specific for a QMC framework, indeed App.~\ref{app4} is specifically oriented to an implementation of the method into a Hartree-Fock or DFT framework.
Next we show how to sample the introduced partition function using  a Langevin dynamics for the wave function parameters, sections~\ref{s_MCsamplingZQ}, and the ion coordinates, section~\ref{s_covlangdyn}.
In section~\ref{s_results} we finally show  some results we have obtained  using this approach for the hydrogen molecule.


\section{Finite temperature electronic partion function} \label{s_upperbound}

We consider the problem to estimate the finite temperature 
partition function of an electronic system with $N$ electrons and $M$ atoms, where we assume in the following that, as discussed in the introduction, 
the ions are  
classical particles, whose coordinates ${\bf R}$ appear just as 
simple parameters in the electronic hamiltonian $H_{\bf R}$ and are confined 
in a finite volume $V$.
Therefore, once the ion positions are fixed, 
 we need to evaluate the electronic partition function:
\begin{equation} \label{trace}
 Z[{\bf R}] = \Tr \exp \lef  -\beta H_{\bf R }  \rr
\end{equation} 
where $\beta =  1 / T $.
Our derivation applies for an Hamiltonian
with a bounded  spectrum defined 
in a finite  Hilbert space with dimension $D$.
For instance in electronic structure calculation one can consider a finite 
dimensional basis of localized orbitals around each atom.
In order to simplify  the notations we can consider standard creation operators 
with canonical commutation rules, spanning the finite single electron basis,
as for a standard lattice Hamiltonian,
namely $ c^{\dag}_i $ for $i=1,\cdots, L$, where for shorthand notations 
$i$ labels also 
the spin, namely $i\le L/2$ ($i>L/2$) refers to spin up(down)-states. 
We consider the generic wavefunction $| \psi \rangle = J \times |SD \rangle$, where:
\begin{eqnarray} \label{formwf}
J &=& \exp( 1/2  \sum_{i,j}  v_{i,j} n_i n_j ) \\ 
|SD \rangle &=& \left[ \prod\limits_{i=1}^N \sum\limits_{j=1}^L  \psi_j^i c^{\dag}_j 
 \right] | 0 \rangle  
\end{eqnarray}
and $n_i= c^{\dag}_i c_i$, for a system of $N$ electrons.
In the continuous limit this wave function is the standard Jastrow-Slater 
one used in quantum Monte 
Carlo in order to describe electron correlation\cite{Jastrow}.
Estensions of this wave function are possible using AGP\cite{Casula2003}, Pfaffian\cite{pfaffian}, backflow\cite{backflow}, and the following considerations apply also for these more 
recent ansatz, because they all contain the Slater determinant $|SD\rangle$ 
in a particular limit.

In all cases, the real variational parameters that define the above 
wave function, 
namely $v_{i,j}$ and $ \psi_j^i$ are compactly denoted by 
$\alpha \equiv \{ \alpha_i \}_{i=1,\ldots,p}$ 
and since all 
physical quantities do not depend on the norm of the wave function, we 
consider the $\alpha-$manifold of states:
\begin{equation}
| \alpha \rangle ={ | \psi \rangle \over \| | \psi  \rangle \| }  
\end{equation} 

The metric in this manifold becomes non trivial as, 
by a straightforward calculation,  
the distance between two states 
$| \alpha \rangle $ and $| \alpha + d \alpha \rangle$ is given by:
\begin{equation}
ds^2 = \| | \alpha + d \alpha \rangle - | \alpha \rangle \|^2= 
d \alpha^i d \alpha^j S_{i,j}  
\end{equation}
where summation over repeated indices is assumed, and  $S$ is a $p \times p$ 
matrix defining the metric tensor of this rather non trivial space, 
described by $p$ independent variational parameters (e.g. a subset of 
 $v_{i,j}$ and $\psi_{ij}$).  
The matrix $S$ can be explicitly evaluated and depends only on 
average first derivatives of the wave function with respect to the 
parameters $\alpha's$:
\begin{equation}
S_{i,j} = { \langle \partial_i \psi |  \partial_j \psi \rangle 
\over \langle \psi | \psi \rangle }  -  
{ \langle \partial_i \psi |  \psi \rangle 
\over \langle \psi | \psi \rangle } { \langle \psi |  \partial_j \psi \rangle 
\over \langle \psi | \psi \rangle }
\end{equation}
It defines a metric as it is strictly positive definite if all 
$p$ variational parameters are independent and therefore its determinant 
$|S| $ is non vanishing.
This matrix turns out to be  exactly the one used in several optimization 
techniques\cite{Umrigaropt,Sorella2007},  
and can be computed also for correlated
 systems by sampling the correlations of the quantities 
$ O_j (x) = {\langle x | \partial_j  \psi \rangle \over  
\langle x |  \psi \rangle } $ 
over the configuration space $\{ x \}$ where electrons have a definite spin 
and positions, namely:
\begin{equation}
S_{i,j} = \langle O_i O_j \rangle - \langle O_i \rangle \langle O_j \rangle 
\end{equation}
where the symbol $\langle \ldots \rangle$ denotes average over a distribution 
$\Pi (x) \propto \langle x | \psi \rangle^2$, that can be sampled by standard 
variational Monte Carlo.  

In Eq.(\ref{trace}), we use  a simple  relation for recasting  
the trace in a finite dimensional Hilbert space as 
an integral of normalized wavefunctions $| c \rangle = \sum\limits_{i=1}^D  x_i | i \rangle $, namely:
\begin{equation} \label{inttrace}
D \int dx^D \delta ( \|x\|-1) \langle c | \exp( -\beta H ) | c \rangle = 
S_D \Tr \exp ( - \beta H )  
\end{equation}  
where $S_D= 2 \pi^{D/2}/\Gamma(D/2)$ is the area of the $D-$dimensional unit sphere.
We note that this simple relation can be used  
to  establish within a  rigorous mathematical framework 
 the finite temperature Lanczos 
method used in Ref.\onlinecite{prelovsek}. In this technique 
finite temperature 
estimates of the partition function are obtained 
with a finite set of randomly generated states $|c\rangle$, 
once it is assumed that $\langle c | \exp( -\beta H ) | c \rangle $, can be 
computed with high accuracy with the Lanczos method.
Indeed this is nothing but evaluating statistically the integral in the 
LHS of Eq.(\ref{inttrace}), and one does not need any further assumption 
to validate the method, apart from the fact that error bars have to be 
computed with standard statistical techniques. 

The simple relation (\ref{inttrace}) 
 can be also extended in the  space $\alpha$ with non trivial metric, 
by using the invariant measure $ d \alpha^p \sqrt{ | S| } $, 
 corresponding to the metric tensor $S$:
\begin{equation} \label{basic}
  { \int d\alpha^p \sqrt{ | S | }   \langle \alpha | \exp ( - \beta H_{\bf R} ) | \alpha  \rangle  \over  Z_S }  = \Tr \exp ( - \beta H_{\bf R} ) 
\end{equation} 
This relation is proven in App.\ref{app1}, provided the 
dimension of the space is large enough, namely contains at least the full space 
of Slater determinant wave functions, 
where the overall constant has 
been obtained by using that $Z[{\bf R}]=D$ for $\beta=0$, as 
the metric normalization $Z_S$ is defined 
as $Z_S={\int d\alpha^p \sqrt{ | S | } \over D} $. 
We emphasize here that the relation (\ref{basic}) is {\em exact} even when 
the dimension of the space $p$ is much smaller than the dimension of 
the Hilbert space. For instance for real Slater determinants the number 
$p< N L$ as they are defined by $N$ orbitals each depending on $L$ 
coefficients (see Eq.\ref{formwf}), whereas the Hilbert space dimension $D$ 
grows exponentially with $L$ and $N$
(See App.\ref{app4} for the parametrization of an arbitrary real Slater Determinant))

In practice the number $p$ of variational parameters  
defining the wave function ansatz 
can be much smaller than that  necessary to span all possible 
Slater determinants.  
In the case $p \ll NL$ we expect that the equation (\ref{basic}) 
is still valid but the trace in the RHS is limited to the largest subspace 
 with dimension $D_s$ spanned by the variational ansatz. Moreover a weak 
dependence on $R$ in $Z_S$ is also expected when a basis dependent on the 
atomic positions is used 
(it is not the case for a plane wave basis for instance).
The calculation can be meaningful also in this case  
after a careful  study of the dependence of the results upon the
 dimension of the basis chosen, as it is common practice in quantum chemistry
calculations.  
In fact, in the limiting case when the one particle basis set 
used to define the orbitals in the Slater Determinant becomes 
 complete the metric normalization $Z_S$ is independent of $R$, 
because any change of basis is equivalent in this limit to a 
mapping $\alpha \to \alpha^\prime$. 
Thus  $Z_S$, being explicitly covariant, is  independent of $R$ and 
can be considered as an irrelevant constant. 
Therefore, within the completeness assumption, following the simple derivation 
 of App.\ref{app2}, 
 we can easily bound the exact electronic partition function 
$Z[{\bf R} ]$, because, due to the convexity of the 
exponential function, the expectation value of  an 
exponential operator over a normalized state $ | \alpha \rangle $ 
satisfies: 
$$ \langle \alpha | \exp ( - \beta H_{\bf R } ) | \alpha \rangle \ge 
\exp ( -\beta   \langle \alpha  |  H_{\bf R} | \alpha \rangle ).$$
This immediately provides a rigorous lower bound $Z_Q$ for the  partition 
function $Z$:
\begin{equation}
Z \ge Z_Q   =  { \int d {\bf R}  \int d\alpha^p \sqrt{ |S|} \exp ( -\beta  \langle \alpha  |  H_{\bf R} | \alpha \rangle ) \over Z_S }  \label{defzq}
\end{equation}
and a corresponding upper bond $F_Q$ for the  free energy $F =-T \ln Z$ :
\begin{equation} 
 F \le  F_Q =  -T \ln Z_{Q}                
\end{equation}

In this way it is evident that $F_Q $ represents 
an improvement to the standard Born-Oppenheimer (BO) approximation.  
In fact in this approximation   
only one state is assumed to contribute to the integral in Eq.(\ref{defzq}), 
namely the lowest energy state of $H_{\bf R}$ within the ansatz given by $|\alpha \rangle $:
\begin{equation}
E_{BO}[{\bf R}]  =  \min_{\alpha} \left\{ \langle \alpha | H_{\bf R} | \alpha  \rangle \right\}
\end{equation} 
Indeed it is clear that $F = \min_{\bf R} \left\{ E_{BO}[\bf R] \right\}$ only at $T=0$, 
and represents a very bad approximation to $F$ as long as the temperature 
is raised, whereas the approximate partition function $F_{Q} $
approaches the correct large temperature limit $-T \ln (D V^{M})$ of the exact 
partition function, while remaining a rigorous upper bound for any $T$. 

In App.\ref{app_comparisonBO} we see in detail a comparison between the approximated partition function $Z_Q$ here introduced, and the exact and  BO ones, 
showing that 
our approximation turns out to be better than the BO one,  
above a temperature $T^*$, that remains meaningful in the thermodynamic limit.

\section{ Monte Carlo sampling of the partition function $Z_Q$ } \label{s_MCsamplingZQ}

In principle the partition function $Z_Q$ can be sampled by almost standard 
Monte Carlo methods, whenever the metric $S$ and the expectation value of the 
energy $H$ over the ansatz $| \alpha \rangle$ are known, for instance within 
the Hartree-Fock theory, namely when $|\alpha\rangle$ represents 
just a simple Slater determinant. 
It is also  possible to replace in $Z_Q$ the expectation value of the 
energy with any DFT functional depending on $|\alpha\rangle$, through the 
corresponding density or gradient, the condition of functional 
minimum being recovered correctly at $T=0$. 
For a discussion about the space of parameters for a Slater determinant wave function, and the introduction of an invariant measure in this space see App.\ref{app4}.

However in the truly correlated case,  
namely when the ansatz $|\alpha \rangle$  differs from a Slater determinant, 
there are  extra complications because  both  the matrix 
$S$ and $\langle \alpha | H_{ \bf R} | \alpha \rangle $ are known only 
within statistical accuracy.
In this case a possible  way to sample the partition function $Z_Q$ and corresponding thermodynamic quantities is to use the penalty method\cite{penalty}, 
introduced some years ago, by using a  cost function
\begin{equation} \label{penalty}
V_P( \alpha , {\bf R} )=\langle \alpha | H_{ \bf R } | \alpha \rangle - {1 \over 2 \beta} \ln | S | 
\end{equation}  
that can be computed statistically with corresponding error bars.

In the following we have chosen a different route, by employing 
a finite temperature  molecular dynamics rather than 
Monte Carlo sampling, because 
recent quantum Monte Carlo packages provide efficient estimates of energy derivatives and ionic forces\cite{Sorella10}.

Our goal is to sample points in the electronic parameter space $\alpha$ distributed according to the probability distribution
defined in Eq.(\ref{defzq}), by using 
first order derivatives of the cost function.
 In the standard Cartesian metric it  is common practice to use a 
Langevin dynamics  for the variables $\{\alpha \}$ and $\{ \bf R \}$, 
 by means of the standard 
first order equation of motions (unit mass is assumed for simplicity)\cite{Attaccalite08}:
\begin{eqnarray}
 \dot {\vec x}  & = -  \partial_{\vec x}  V  + \vec \eta 
\end{eqnarray}
where $\vec x$ is
a covariant vector in a finite dimensional 
euclidean space, whereas $ \partial_{\vec x}  V(x)$
 is the derivative (force) of 
a potential $V$.
By means of this equation it is well known that it is possible to sample 
the equilibrium distribution $W_{eq} (x) = \exp ( - \beta V(x) ) $ provided 
we satisfy the fluctuation dissipation theorem given by:
\begin{equation} \label{fdeq}
 \langle  \eta_i(t)  \eta_j(t^\prime)  \rangle  =\delta(t-t^\prime) \delta_{i,j}  { 2  \over \beta }
\end{equation}
Now we suppose to change the reference coordinate system by means of a generic
 transformation of variables
$ x \to \alpha$ (an $N-$ dimensional non linear mapping as in general 
relativity).
Be the Jacobian of the transform given by the matrix $L$:
\begin{equation}
L_{i,j} = \partial_{x_j} \alpha_i ( \vec x) 
\end{equation}
The Langevin equation in this new reference can be easily obtained:

\begin{eqnarray} 
\label{eq_1ord}
\dot {\vec \alpha } = - S^{-1} { \partial V \over \partial \vec \alpha } + L \vec \eta 
\end{eqnarray}
where $ S^{-1} = L L^\dag $, 
and the equation (\ref{fdeq}) that defines the fluctuation 
dissipation theorem remains unchanged.

The Eq.(\ref{eq_1ord})  is covariant if we just replace the matrix $S$ with the matrix defining the metric in a generic curved space:
\begin{equation}
ds^2 = S_{i,j} d\alpha_i d \alpha_j 
\end{equation}
where, as usual, in this formalism repeated indices are assumed summed.
Indeed after the given transformation the above metric tensor transforms 
as:
\begin{equation}
 S \to (L^{\dag})^{-1} S L^{-1}
\end{equation}
that, as it should,  leaves unchanged the covariant first order Langevin 
equation (\ref{eq_1ord}).

Thus,  from the above equation, 
 we obtain the desired result with the matrix $L$ given by 
any solution of the matrix equation:
 $$ S^{-1} = L L^\dag. $$ 


Unfortunately Eq. (\ref{eq_1ord}) looks a bit complicated when it 
is  discretized in times $t_n = \Delta n$, because the integral of the 
random noise depends explicitly on the curvature of the non linear space 
by means of the matrix $L$, 
and the resulting integration is not univocally defined, simply because 
the solution $S^{-1}=L L^\dag$ is not unique, since $S^{-1}$ remains 
unchanged under the substitution $L \to L U$, where $U$ is an arbitrary 
unitary matrix.
In order to remove this arbitrariness, according to Risken\cite{risken}, 
we can work out the integral of the 
equation of motion in a small time interval  of length $\Delta$, 
by requiring also that the corresponding Markov process:
\begin{eqnarray}
\label{eq_a}
  \alpha(t_{n+1})^i &= &\alpha (t_n)^i  -\Delta \left[ S^{-1}(t_n)   \partial_{\vec \alpha} \left( V - {1\over 2 \beta } \ln {\rm Det} S \right)(t_n) \right]^i  \nonumber  \\
&+& {1\over 2 } \sum\limits_{k} \partial_{\alpha_k} D_{i,k} + y_n^i \nonumber \\
\langle y_n^i y_n^j \rangle &=& D_{i,j} = {2 \Delta \over \beta } S^{-1}_{i,j} (t_n) 
\end{eqnarray}
 has the correct equilibrium 
distribution for $\Delta \to 0$:
\begin{equation} \label{equilibrium}
W_{eq} ( \alpha ) \propto \sqrt{ {\rm Det} S } \exp ( - \beta V(\alpha) ) 
\end{equation}
In fact it is possible to show that, only with the above definition of 
the drift term, the associated and univocally defined 
 Fokker-Planck equation for the probability distribution $W(\alpha,t)$ reads for $\Delta \to 0$:
\begin{eqnarray}
\partial_t W(\alpha,t) &=& \sum\limits_{ j} \partial_j 
\left\{ \sum\limits_i 
{ 1\over \beta }  S^{-1}_{j,i} \partial_i W(\alpha,t) \right.  \\ 
&+& \left.  W (\alpha ,t) \left[ S^{-1} \partial_{\vec \alpha} \left (V(\alpha)- {1 \over 2 \beta } \ln {\rm Det} S \right)  \right]^j \right\} \nonumber
\end{eqnarray}
which has the equilibrium distribution $W_{eq}(\alpha)$ satisfying:
\begin{equation}
\sum_i \frac{1}{\beta} S^{-1}_{j,i} \partial_i W_{eq}(\alpha) + W_{eq} (\alpha) \sum_i S^{-1}_{j,i} 
\partial_i \left( V -{1 \over 2 \beta } \ln {\rm Det} S \right)  =0
\end{equation}
Indeed, by multiplying both sides of the equations by $S_{k,j}$ and 
summing over $j$, we obtain the standard equation for the 
equilibrium distribution $\sqrt{|S|}  \exp( - \beta V)$.

\section{Covariant Langevin dynamics for ions and electrons}\label{s_covlangdyn}

We want to implement the above formalism in an ab-initio molecular dynamics (MD) at finite temperature dealing with electrons and ions within the same formalism, similarly to what was done in the pioneer work by R. Car and M. Parrinello\cite{carparinello}.
In the following we will show how the ionic motion can be quite naturally included in the above scheme.
In fact what we obtained before does not hold only for the electronic parameters, but for a generic set of parameters which appear 
in a variational wavefunction. 
The ionic positions $\bf R$ can thus be thought as complementary parameters.
The inclusion of this kind of parameters in the above formalism is straightforward: if $M$ is the number of atoms, then $S$ becomes a $(p + 3M) \times( p + 3M) $ block-diagonal matrix. The mixed elements $S_{\{ \alpha\},\{\bf R\}}$ are always zero since wave functions characterized by different sets of atomic positions are orthogonal.
Moreover, since the ionic positions $\bf R$ belong to the real space, the corresponding metric is the Cartesian one, and is defined by a diagonal matrix 
$S(R_l, R_r)= S_N \delta_{l,r}$ among all the ion components.  
We can esplicitly write down the complete set of equations for both the atomic and electronic parameters.
For the ionic positions we use
\begin{eqnarray}
\label{eq_r}
R(t_{n+1})^l &=& R(t_{n})^l + \Delta_N ~ F^l(t_{n},\{\alpha(t_{n}\}) + \chi^l_n \nonumber \\
\langle \chi_n^l \chi_n^r \rangle &=& {2 \Delta_N \over \beta}  ~ \delta_{l,r}
\end{eqnarray}
with $l,r=1,\cdots, 3M$ and $F^l$ being the force acting on the $l$-th ionic cartesian coordinate,
while for the electronic variables Eq. (\ref{eq_a}) holds
with $i,j=1,\cdots,p$ and
where $-\partial_{\vec \alpha} V$ is the force acting on the parameters $\alpha$, i.e, the gradient of the total electronic energy $V$ evaluated at fixed $\bf R$ with respect to these parameters. 

Notice also that the time discretization corresponding to the ionic dynamics 
is defined by the arbitrary constant $S_N$ appearing in  the extended 
metric tensor defined before,  namely $\Delta_N= \Delta S_N^{-1}$.
It is clear therefore that the relative speed between electron and ion dynamics 
can be tuned to optimize efficiency, exactly as in Car-Parrinello ab-initio
 molecular dynamics. We emphasize here that in the limit $\Delta, \Delta_N \to 0$ consistent results are obtained because the equilibrium distribution 
(\ref{equilibrium}) remains unaffected by the choice of $S_N$.


\section{Results and Discussion}\label{s_results}

Once we set up the discretized equations (\ref{eq_a},\ref{eq_r}) we can test the above formalism in a simple but realistic case.
We are going to study the $H_2$ molecule, looking at the temperature behavior of the total energy $E$ and the bond distance $r$ between the two hydrogen atoms. We start with this simple system because the above quantities can be easily 
computed, providing therefore useful benchmarks for our technique.

According to App.\ref{app_comparisonBO}  
the distribution sampled by means of this covariant Langevin dynamics (CLD) 
represents an improvement of the BO only above a temperature $T^*$.
At $T=0$  our approximate free energy $F_Q$ coincides with the BO one 
$F_{BO}$, 
but as soon as  $T>0$ the $F_{BO}$ becomes better for $T \le T^*$.

If the temperature is much lower than the electronic 
gap  the BO approximation  should be essentially exact
and can be easily obtained from the potential energy surface (PES) 
$v(r)$ of the $H_2$ molecule.

In the following we are going to show  that, in this simple system, we cannot 
distinguish  the correct  BO low temperature behavior and the 
one implied by our approximate technique, clearly   
indicating that $T^*$ should be almost negligible for this system.

\begin{figure}[h]
\begin{center}
\includegraphics[scale=0.33]{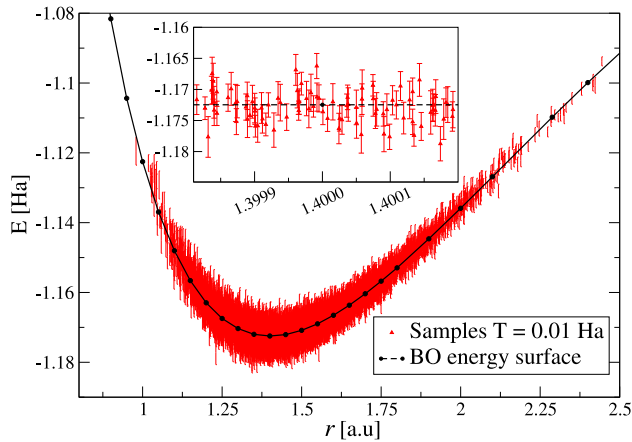}
\caption{Black line: Total energy $E$ as a function of the bond length $r$ obtained by minimizing the energy of our variational wavefunction for fixed $r$; 
in doing this we act only on those parameters $\{\alpha \}$ which are kept free to evolve in the dynamics (\ref{eq_a}). 
Red points: Energy with error bars of configurations sampled in a dynamics (\ref{eq_a},\ref{eq_r}) with $T=0.01 ~Ha$. The PES is correctly followed during the simulation. In the inset a little region around the minimum at $r=1.40~ a.u.$ is enlarged.}
\label{f_pesh2}
\end{center}
\end{figure}

To proceed further we need now to specify what type of correlated variational wavefunction (\ref{formwf}) we adopt in all the following calculations, and 
its dependence on the two electronic positions $\vec r_1$ and $\vec r_2$.
In the singlet state the orbital function $f(\vec r_1, \vec r_2)$ 
is symmetric and positive and is parametrized here as a product of two 
factors $f(\vec r_1,\vec r_2) = f_0(\vec r_1, \vec r_2) 
\times \exp (J( \vec r_1, \vec r_2)) $, where $f_0$ is taken fixed and allows
 to satisfy the electron-electron and electron-ion cusp conditions, whereas 
\begin{equation} \label{defj}
J= \sum\limits_{i,j}   \lambda_{i,j} \phi_i( \vec r_1) \phi_j (\vec r_2)
\end{equation} 
is cusp free and is expanded systematically in a basis of atomic 
orbitals centered on each atom containing up to  
3s and 1p  gaussian functions and a constant one $\phi_0=1$.  This amounts  
 to  $p=65$ independent variational parameters for the 
symmetric matrix $\lambda_{i,j}$.
The exponent of the gaussians are kept fixed during our simulations.
Despite this limitation in the choice of the basis this is acceptable for 
the $H_2$ molecule in a physically relevant range of 
distances between the atoms, as it is shown in Fig.(\ref{f_pesh2}).
 
The chosen variational ansatz is particularly useful for evaluating 
the complicated terms in (\ref{eq_a}), i.e. the \emph{drift-diffusion} ones which depend linearly from the temperature and require the knowledge of the derivative of the matrix $S$. This is indeed simpler for the  parameters $\lambda_{i,j}$ which appear in a linear fashion in  the exponential factor $J$ in Eq.(\ref{defj}). 
The first step is thus to construct the PES of the molecule ( Fig.\ref{f_pesh2}). In this way we not only acquire the key information for the numerically exact evaluation of the BO observables, but we also check that our choice of the free variational parameters in the wave function allows us  to recover the well known PES for this molecule\cite{kolos,Pachucki2010}.

\begin{figure}[h]
\begin{center}
\includegraphics[scale=0.33]{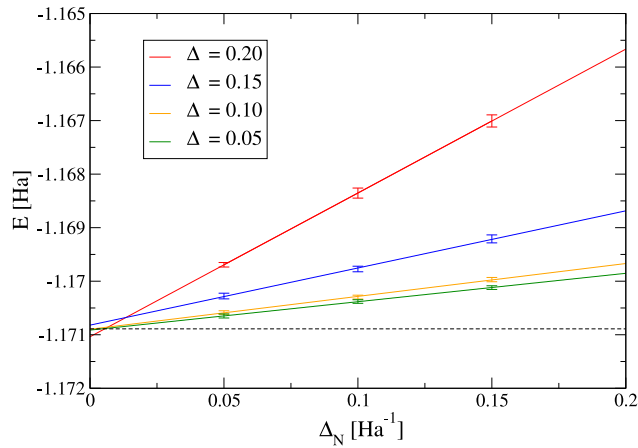}
\caption{Time averages of the total energy $E$ at $T=0.003~ Ha$ as a function of $\Delta_N$ for $4$ values of $\Delta$. All the series converge roughly to the same value which is also the expected one (horizontal dashed line) obtained with eq. (\ref{e_can_av}), simplifying the second extrapolation $\Delta \to 0$. }
\label{f_extrap}
\end{center}
\end{figure}

Canonical averages of an observable $O(r)$ can be obtained by computing numerically the one dimensional (conditionally convergent) integrals 
\begin{equation}
\label{e_can_av}
 \hat{O} = {\int  dr ~r^2~ O(r) \exp \lef -\beta v(r) \rr \over \int  dr ~r^2 \exp \lef -\beta v(r) \rr}
\end{equation}
On the other hand we can compute $\hat{O}$ as a time average on the Langevin dynamics (\ref{eq_a},\ref{eq_r}) for sufficient low $T$. 
The extrapolation $\Delta \to 0$ involving the discretized time steps is performed in the order $\Delta_N \to 0,\Delta \to 0 $. 
It is observed (see Fig.\ref{f_extrap}) that the $\Delta_N$ dependence of the time averages of the quantities is linear for fixed $\Delta$, a property useful in the extrapolation.

\begin{figure}[h]
\begin{center}
\includegraphics[scale=0.33]{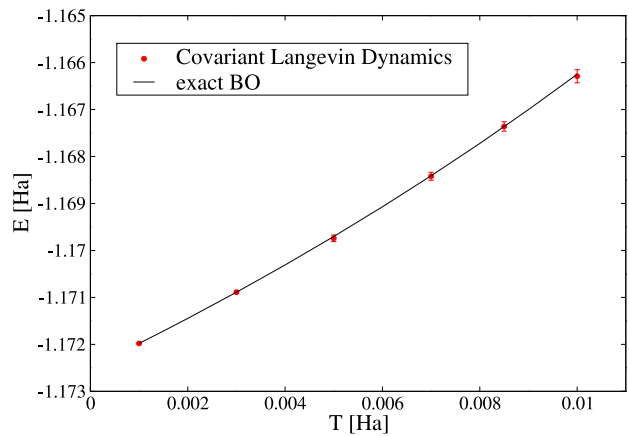}
\caption{Total energy $E$ as a function of temperature. The range of temperature is well below the electronic gap $\sim 0.17~Ha$ (see Fig.\ref{f_pesh2}) so the expected exact value is the BO one evaluated by eq. (\ref{e_can_av}) (black line). Red points are obtained by integrating the coupled equations (\ref{eq_a},\ref{eq_r}). Data are in agreement with the predicted values.}
\label{f_eT}
\end{center}
\end{figure}

\begin{figure}[h]
\begin{center}
\includegraphics[scale=0.33]{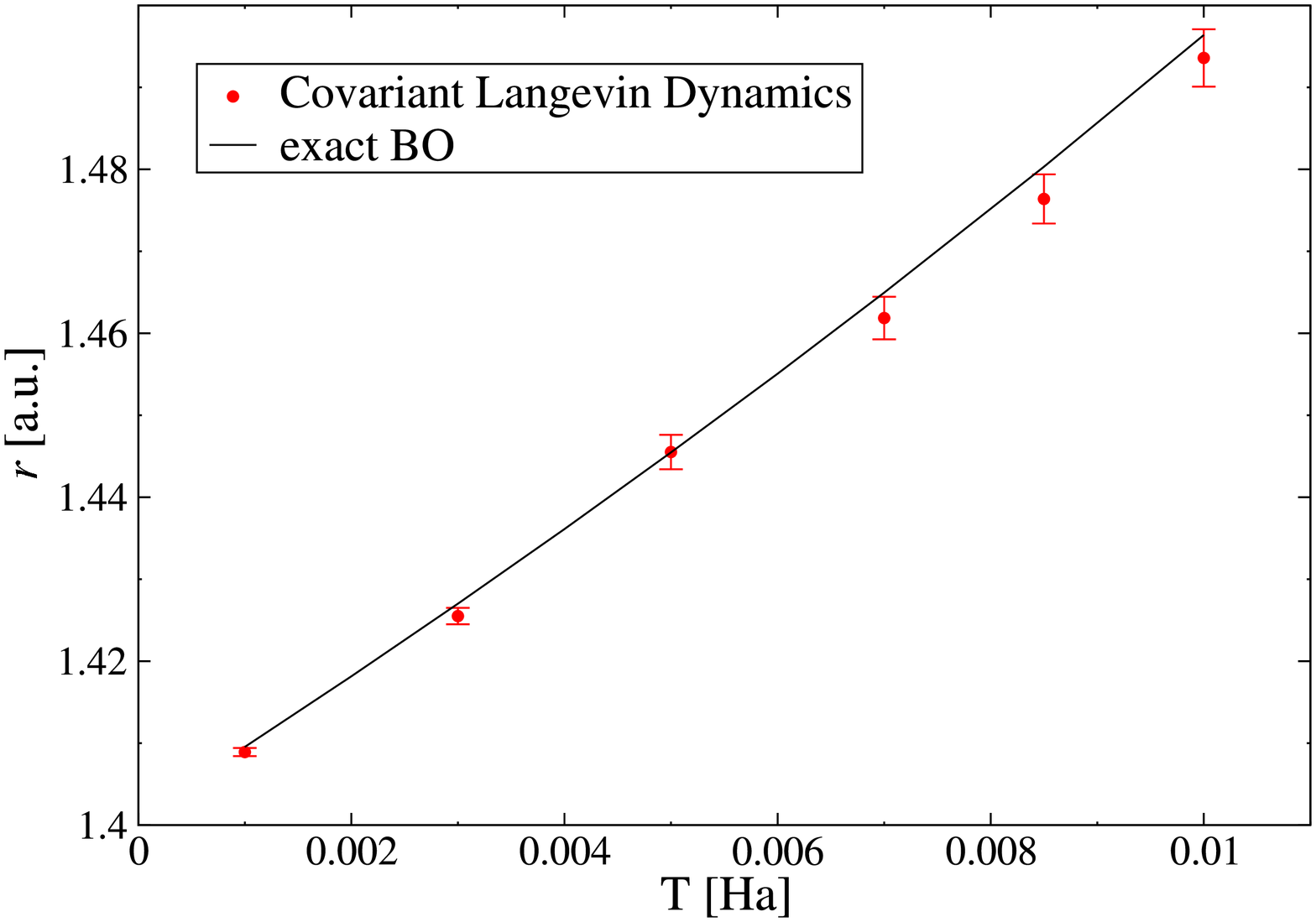}
\caption{Bond length $r$ as a function of temperature. Even for this observable the Langevin dynamics (red points) give values compatible with the expected ones (black line).}
\label{f_rT}
\end{center}
\end{figure}

Finally we show our results for the total energy and the bond distance at varius temperatures in the range between $0.001 \div 0.01 ~ Ha$, i.e. from room temperature to $\sim 3000~ K$.
The forces acting on the parameters and on the ions, as well as the matrix $S$ are evaluated by a short QMC run at each iteration of the dynamics.
In Fig.(\ref{f_eT}) and in Fig.(\ref{f_rT}) we  show the outcome of our covariant Langevin dynamics simulations. 

We see that our Langevin dynamics gives result in very good agreement with the expected BO values.
 We stress once again that this dynamics does not require an electronic minimization at each ionic move, realizing an impressive gain from the point of view of the computational cost.
\begin{figure}[h]
\begin{center}
\includegraphics[scale=0.33]{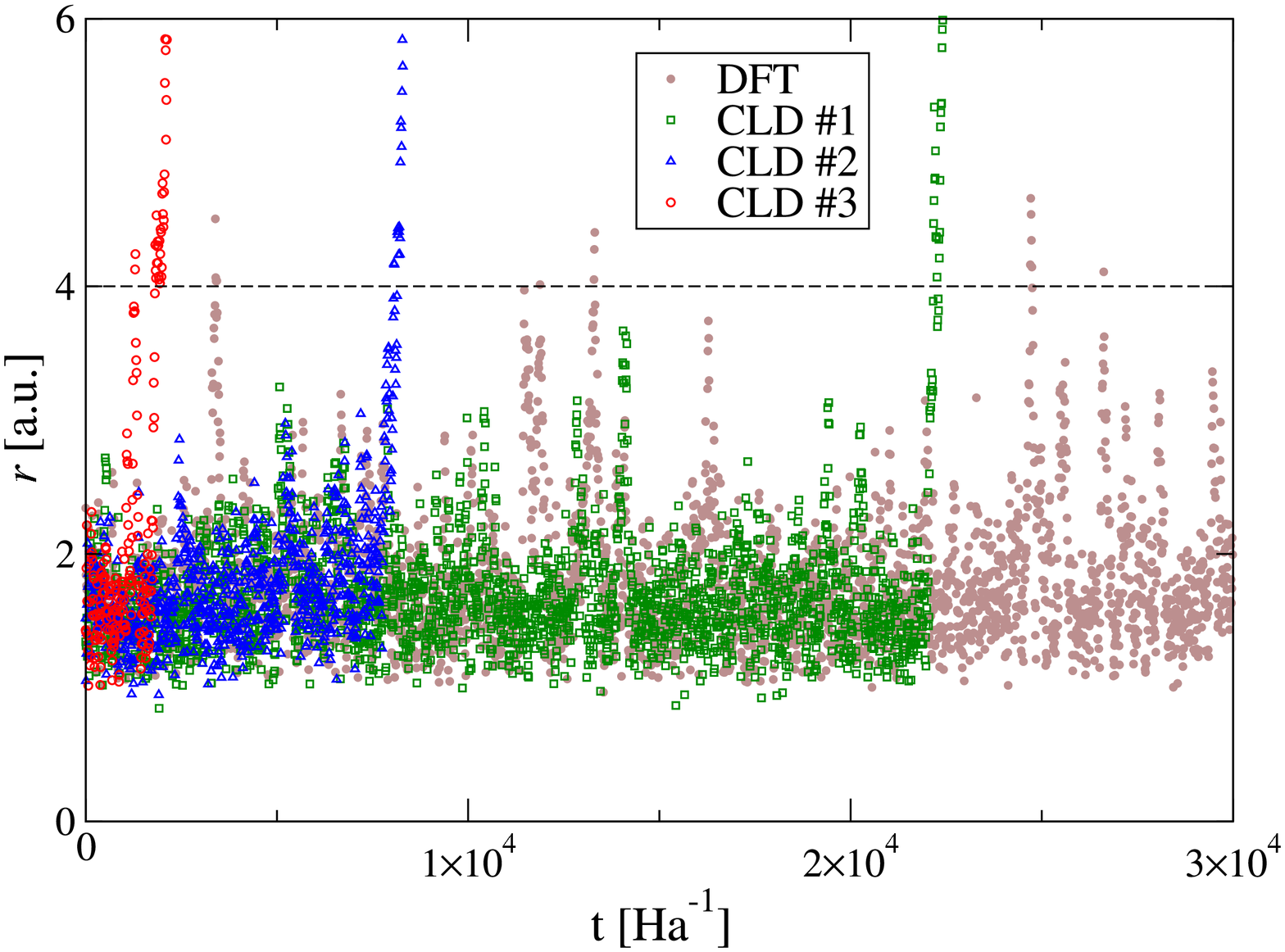}
\caption{Bond length $r$ as a function of simulation time at a temperature of $T=8000~K$. Coloured points (red, green and blue) correspond to simulation performed with the dynamics presented in this work, while the grey solid ones are obtained with a DFT- Langevin BOMD. The time step used in the integration of the equations is $\Delta_N = 0.1~ Ha^{-1}$ and points are plotted every $10$ iterations. The dashed line indicates the  distance $r^*$ such that the energy gap between the ground state PES and the first excited one becomes smaller than $T$. All the CLD trajectories show escape events while the DFT one describes a stable molecular configuration up to $20\times10^4~Ha^{-1}$ of simulation time (not shown).}
\label{f_dissoc}
\end{center}
\end{figure}
On the other hand, this kind of dynamics should behave differently with respect to the standard BO-MD one when the temperature is raised and for $T>T^*$ 
should be more realistic, because corresponding to a more accurate upper 
bound of the exact free energy $F$.
In figures (\ref{f_eT},\ref{f_rT}) we limit the study of the average energy and bond length in a range of temperatures smaller than $ ~3000~K$ because, above this value, first dissociation events start to appear during the simulations. This temperature is in good qualitative agreement with low pressures experiments\cite{langmuir}.
 Roughly speaking the dissociation probability depends on the ratio between the thermal energy $T$ and the depth of the free energy well $\Delta U$ through the Boltzmann weight\cite{hanggi} $\exp (-\Delta U / T)$ within  
 the assumption that excited electronic eigenstates are well-separated in energy from the ground state.
There are instead examples\cite{butler} in which BO approximation breaks down, 
particularly near the transition state of a chemical reaction. 
In fact, as the reaction coordinate $r$ increases, the energy gap  between the ground state and the first (antibonding) excited state becomes smaller\cite{kolos},
for example when $r > 4~a.u.$ this quantity becomes smaller than  $8000K$.
Therefore large fluctuations in the bond length, certainly occurring 
at large temperatures,  are in principle not well described under a BO scheme.
Since by definition, an atomic dissociation requires to sample correctly 
events with large 
 $r$,  we expect to find differences between the standard BO-MD and the dynamics generated by (\ref{eq_a},\ref{eq_r}), at large enough temperatures.
In  Fig.\ref{f_dissoc}, we observe that the probability of dissociation is enhanced in our dynamics, which can take implicitely into account also the effective repulsion due to the antibonding state. As expected, this is in sharp contrast 
with a DFT-BO dynamics obtained using the {\sc Quantum ESPRESSO} package\cite{espresso,pseudo}.
In the latter dynamics large fluctuations in $r$ do not lead to dissociation, as is partially shown in fig. (\ref{f_dissoc}). 
Indeed no escape event occurs within DFT BO-MD,  
even for a long time dynamics.
Moreover in order to compensate the well known overbinding error\cite{pippah2} of the local density approximation (LDA), we have 
increased the temperature by a factor proportional to the LDA energy 
barrier ($0.2415 Ha$),
and observed no qualitative changes in the trajectories, always confined 
around the minimum energy value.   
It is clear therefore that, quite generally, the BO-MD greatly  
underestimate the evaluation of the reaction rate if, for instance,  a \emph{mean first-passage time}\cite{hanggi} analysis is performed.

 
\section{Conclusions}
In this paper we have introduced a new promising approach to deal 
with finite temperature simulations of electronic systems.
The approach is general and, as we have emphasized in the introduction,
can be easily extended to several branches of the electronic simulations,
from ab-initio finite temperature  simulation of realistic systems  
based on Hartree-Fock, DFT or quantum Monte Carlo methods, 
to finite temperature simulations of strongly correlated Hamiltonians 
defined on a lattice.
In particular this technique allows us to improve systematically the Born-Oppenheimer approximation in a temperature range where the quantum effects on atoms 
are neglegible. 
In principle also these quantum effects can be dealt in a simple 
way. To this purpose it is enough to define a quantum ansatz $|\alpha\rangle$
describing electrons and ion coordinates quantum mechanically, including in 
$\{ \alpha \}$ also  
variational parameters  corresponding to the atomic wave function 
$\Phi (\bf R)$, 
for instance described by gaussians centered around the average atomic
positions.
In that case the same derivation holds as electrons and ions can be dealt 
in the same footing, the metric matrix $S$ will have non trival 
off diagonal elements between electronic and atomic variational parameters.

Although our first application is limited to the simple $H_2$ molecule 
with classical atomic coordinates,
this extremely simple 
example already shows that it is possible to catch some 
qualitatively new behavior, that is not possible to describe with the conventional BO approximation. 
Namely at large enough temperature the molecule can dissociate due to 
non adiabatic effects.

We plan to extend our method to larger and more complex 
realistic systems including also quantum effects for atoms.
Unfortunately,  so far we have encountered a  difficulty 
to compute in an efficient way 
the metric tensor $S$ and its derivatives for a generic correlated 
wave function. For this reason, at present, it looks that the penalty method\cite{penalty}  could be a more realistic 
possibility for extending our technique, because the penalty method 
does not require the evaluation
of the derivatives of the metric tensor.  
Apart for this technical issue
there are many open problems that can be tackled with this 
new technique. For instance one would like to know the magnetic transition temperature 
of a piece of material. Without taking into 
account the electronic entropy this is not possible for most 
electronic ab-initio methods, but,  by 
applying our technique, a reasonable estimate can be easily obtained.
In lattice models, an old standing problem is for instance the extension 
of the Gutzwiller variational ansatz to finite temperature calculations.
Within the variational Monte Carlo 
 it has been established that 
the Hubbard model for $U/t$ large enough should be superconducting 
with a d-wave order parameter.
However it is not possible  to predict within the same ansatz 
the much more interesting 
superconducting temperature and how it depends on the various details of the model, such as doping and the value of the Coulomb repulsion $U/t$.
In our formulation what can be done at zero temperature can be readily extended to finite temperature and the evaluation of the critical temperature should be 
straightforward, likewise a standard (but much more accurate because including 
electron correlation) mean field theory at finite temperature.

\acknowledgements

We acknowledge useful discussions with A. Laio and S. De Gironcoli.
Computational resources  were supported by  CINECA Bologna.

\appendix

\section{ Proof of the integral formula of Eq.(\ref{basic}) }
\label{app1}
In this appendix we use known results of differential geometry in 
Riemann spaces\cite{helgason} with tensor metric $S$.
In order to prove Eq.(\ref{basic}), it is enough to consider the 
complete basis:
\begin{equation}
|i \rangle = \prod\limits_{n=1}^N  c^{\dag}_{l_i(n)} |0 \rangle 
\end{equation}
where $l_i(n)$ is an arbitrary choice of $N$ different integers among 
the $L$ possibilities, that defines the Hilbert of  space of $N$ 
fermions containing $D= { L \choose N}$ independent states.
It is simple to realize that it is enough to prove that, 
given two arbitrary states $| i \rangle $ and $ | j \rangle $, we have:
\begin{equation} \label{statement}
O_{i,j} = \int d\alpha^p  \sqrt{ | S | } \langle i | \alpha \rangle \langle \alpha | j \rangle = C \delta_{i,j}
\end{equation} 
where $C$ is an overall constant.
Indeed, by assuming that the above equation holds, we can insert in Eq.(\ref{basic}) 
the completeness $I \sum_i |i \langle \rangle i | $ in both the bra and the ket 
of numerator in Eq.(\ref{basic}) and obtain:
\begin{eqnarray} \label{basic1}
  & \int d\alpha^p \sqrt{ | S | }   \langle \alpha | \exp ( - \beta H_{\bf R} ) | \alpha  \rangle  = &\nonumber \\
& \sum\limits_{i,j}   \langle j | \exp ( - \beta H_{\bf R} ) | i \rangle 
 \int d\alpha^p \sqrt{ | S | }  \langle \alpha | j \rangle \langle i | \alpha \rangle =& \nonumber \\
   &C  \Tr \exp ( - \beta H_{\bf R} ) & \nonumber \\ 
\end{eqnarray}
which easily proves Eq.(\ref{basic}).

In order to establish Eq.(\ref{statement}) 
we can consider the group of transformations $ \alpha \to \alpha'$ 
that leaves unchanged the  metric tensor $S$ defined by:
\begin{equation}
U | \alpha \rangle = | \alpha^\prime \rangle 
\end{equation}
where $U$ is a unitary matrix that maps any variational ansatz $\alpha$ 
to a new variational ansatz $\alpha^\prime$ of the form defined 
in Eq.(\ref{formwf}).
To this purpose it is enough to consider the unitary tranformations defined 
by:
\begin{eqnarray}
U_l c^{\dag}_m U_l^{\dag} &=& ( 1  -2 \delta_{l,m} )
 c^{\dag}_m , \; U_l= \exp( i \pi c^{\dag}_l c_l ) \\
U_P c^{\dag}_l U^{\dag}_P &=& c^{\dag}_{p(l)}  
\end{eqnarray}  
where $p(l)$ is an arbitrary permutation of the $L$ indices. 
All the above transformation are real 
and unitary and therefore conserve the distance 
between two arbitrary vectors, implying that the metric $ds^2$ remains 
unchanged under all these transformations, when applied to any 
arbitrary state of the ansatz:
\begin{equation}
ds^2 = S_{i,j}( \alpha) d\alpha^i d\alpha^j = S_{i,j} ( \alpha^\prime) 
d \alpha^{\prime i} d\alpha^{\prime j}
\end{equation}
In differential geometry these transformations are called isometries, 
and represent the basis for the classification of symmetric Riemann spaces.
In this context they are important to prove the main statement of this 
appendix.
Indeed we can consider any isometry as a change of variable in the integral 
and obtain that (since the integration variables are dummy variables we 
can use $\alpha$ in place of $\alpha^\prime$):
\begin{equation} \label{basic2}
O_{i,j} = \int d\alpha^p \sqrt{ |S| }  \langle  i | U^{\dag} | \alpha \rangle 
 \langle \alpha | U |  j \rangle  
\end{equation}
Now since the set of states is complete the matrix elements $O_{i,j}$ 
define univocally an operator in the given $D-$ dimensional Hilbert space.
Therefore by applying the relation (\ref{basic2}) for all isometries 
$U_l$ for $l=1,\cdots, L$, we obtain that this operator $O$ commutes with 
all fermion occupation  number $n_l$ and therefore has to be diagonal,
namely $O_{i,j}= C_i \delta_{i,j}$.
On the other hand we can apply Eq.(\ref{basic2}) for an arbitrary 
unitary permutation $U_P$, that is able to connect any state $i$ of the 
Hilbert space to any other one $|j \rangle$, namely $U_P|i\rangle=|j\rangle$.
Thus it easily follows that:
\begin{eqnarray}
O_{i,i} &=& \int d\alpha^p \sqrt{ |S| } \langle  i | U^{\dag}_P | \alpha \rangle
 \langle \alpha | U_P |  i \rangle \\
& = & \int d\alpha^p \sqrt{ |S| }  \langle  j | \alpha \rangle 
 \langle \alpha |  j \rangle =O_{j,j}
\end{eqnarray} 
implying that $O_{i,i}=C_i$ does not depend on $i$, and this concludes the 
proof of this appendix.

\section{Proof of the upper bound for normalized states} 
\label{app2}
The expectation value of an operator $O$ 
over a normalized state $\alpha$  is 
equivalent to average  $ \langle \psi_i | O | \psi_i \rangle $ 
over the distribution $ p_i = \langle \psi_i | \alpha \rangle^2$ over the eigenstates $ \psi_i$ of the operator 
$O$. In fact it immediately follows that $ 0 \le  p_i \le 1$ and 
that $\sum_i p_i =1$. 
Since  for any distribution $p_i$ and convex function $f$, 
it is well known that, from Jensen's inequality, we have:
\begin{equation} \label{bound1}
 \left< f(H) \right> \ge f(\left<H\right>)  
\end{equation}
where the symbol $\left<O\right>$ means averaging over the distribution $p_i$ of the 
operator $O$,
namely $ \left< O \right> =  \sum_i p_i \langle \psi_i | O | \psi_i  \rangle $.  
Since the operator $H$ is Hermitian, $H$ and $f(H)$ are diagonalized by 
the same eigenvectors, and therefore the distribution $p_i$ is the same 
for both operators and relation (\ref{bound1}) simply follows from 
the convexity of $f$.
Then by using the convexity of the function  $f(x) = \exp( - x/T) $, 
by applying the above consideration to the operator $O = f( H)$, 
we obtain: 
\begin{equation}
\langle \alpha  | \exp ( - H/T ) | \alpha \rangle  \ge 
\exp( \langle \alpha | - H/T  | \alpha \rangle )  
\end{equation}
which concludes the proof of this appendix.


\section{Approximate partition function $Z_Q$ versus exact and Born-Oppenheimer partition functions} \label{app_comparisonBO}

In this appendix we want to investigate the nature of the approximation of the partition function $Z_Q$ defined in (\ref{defzq}) and used in this work. In order to do this we will compare the approximate partition function $Z_Q$ with the exact $Z$ and the approximate Born-Oppenheimer $Z_{BO}$, in the general 
 case when we use $p < D$ variational parameters in the normalized wave 
function ansatz $|\alpha \rangle$. To simplify the notations we avoid to use 
the dependence on the atomic positions $R$. 
We assume that the ground state energy $E_0$ is non degenerate and
all the eigenvalues $|E_i| \le B$, namely the spectrum is bounded and $B$, as well as the maximum gap $\Delta = {\rm Max}_i E_i-E_0$, grows at most linearly  with the number $N$ of electrons.
These assumptions are commonly satisfied by physical Hamiltonians
of interacting fermions. 

Within these assumptions, we will see that $Z_Q(T)$ is an approximation for $Z(T)$ better than $Z_{BO}(T)$ as long as the temperature $T$ is larger than a crossover  temperature $T^*< \bar T$  where $\bar T$ remains finite for $N\to \infty$.

As mentioned, we assume to know a complete orthonormal set 
$\left\{ |i\rangle \right\}_{i=0,\ldots,D-1}$ of eigenstates of the hamiltonian $H$ that operates in a $D$-dimensional Hilbert space.
This implies that at a given temperature $T$ the exact partition function is:
\begin{equation} \label{Z_app3}
 Z(T) = \sum_{i=0}^{D-1}  e^{- E_i /T} \\
\end{equation}
whereas the BO partition function is:
\begin{equation} \label{ZBO_app3}
Z_{BO}(T) = \exp( -  E_V /T)
\end{equation}
where $E_V = {\rm Min}_{\alpha } \langle  \alpha | H |  \alpha \rangle $ and
the approximate partition function $Z_Q$  is given in Eq.(\ref{defzq}).
We remind  that we have already proven, using the convexity of the exponential function, that the relation:
\begin{equation} \label{rel_Z_ZQ_app3}
 Z(T) \ge  Z_Q(T) 
\end{equation} 
holds for every $T$, and obviously $Z(T) \ge Z_{BO}(T)$.

In order to identify  the  more accurate approximate partition function,
namely the one with the larger bound for $Z(T)$ 
we consider  the ratio between the $Z_Q$ and $Z_{BO}$: 
\begin{equation} \label{zetaq_app3}
\zeta_Q(T) \equiv {Z_Q(T) \over Z_{BO}(T)} 
\end{equation}
Since $Z_Q(T)$ is essentially a classical partition function over $p$ 
variables, the equipartition theorem immediately implies that:
\begin{equation}
Z_Q(T) \propto Z_{BO} (T)   T^{p/2}
\end{equation}
Thus the BO approximation is better at low enough temperature, and, 
our low temperature free energy $F_Q = E_V - p/2 T \ln T $ is expected to 
be a very bad approximation of the quantum free energy especially when 
$p$ is very large, just because classical and quantum free energy 
differ substantially at very low temperatures. 

The above consideration could lead to the disappointing conclusion that 
$\zeta_Q(T) >1$, namely $F_Q(T) \le E_V$, only for very high temperatures.

However we can easily find a lower bound for $\zeta_Q(T)$ by using 
that the spectrum is bounded, as assumed at the beginning of this appendix:

\begin{equation}
\zeta_Q( T)   = D { \int d\alpha^p \sqrt{ |S|} \exp ( -{\langle \alpha  |  H-E_V| \alpha \rangle \over T}  ) \over \int d\alpha^p \sqrt{ |S|} } 
\ge D \exp ( - \Delta /T )  
\end{equation}  
When the above bound is larger than one, 
$\zeta_Q(T)$ is certainly larger than one, implying $F_Q \le F_{BO}$. 
This occurs for $T\ge \bar T$
, where $\bar T$ is easily  determined by $\bar T = \Delta/\ln D$.
Hence  in the thermodynamic limit there exists a 
finite crossover temperature $T^*$, as $\Delta /\ln D$ remains finite 
for $N\to \infty$, according to our assumptions.

\section{Slater determinants and symmetric Riemann spaces} 
\label{app4}

We consider the space ${\cal M}$ 
of normalized Slater determinants in a finite 
dimensional Hilbert space ${\cal H}$ where fermions can occupy $L$ different one particle 
states, denoted by conventional creation operators  $c^{\dag}_i$.
A Slater determinant with $N$ electrons 
 can be formally written in second 
quantization notations by means of $N\times L$ real numbers $\psi^i_j$:
\begin{equation} \label{slater}
 | \psi \rangle = \prod\limits_{i=1}^N \sum\limits_{j=1}^L \psi_{i,j} c^{\dag}_j  | 0 \rangle  
\end{equation}  
However all the variables of the matrix $\psi$ 
are highly redundant because, as well known, 
the Slater determinant after the linear  
transformations $ \psi\to  \hat L \psi $ 
is multiplied by a constant $| \psi \rangle \to | \hat L | |\psi \rangle$, where$\hat L$ is an arbitrary  $N\times N$ matrix and $|\hat L|$ its 
determinant.  
It is clear that, in order to define a Slater determinant with 
unit norm we can consider  one  constraint $ \langle \psi | \psi \rangle = 
| \psi \psi^{\dag} |=1$
over the $N L$ variables defining the $N\times L$ matrix 
$\psi$, amounting therefore to $NL-1$ independent real variables. By the above discussion,  the wavefunction $| \psi\rangle$ is left invariant for all matrix transformation $ \psi \to \hat L \psi$  with $ | \hat L | =1$,  
defining $N^2-1$ independent variables 
for $\hat L$.
Thus, it follows that $| \psi \rangle$ can be parametrized by 
 $ (N L-1)- (N^2-1) =  N (L-N)$ 
 independent real variables. 
In a more rigourous  mathematical formalism, 
by neglecting an immaterial overall sign $\pm 1$ in the definition of $\psi$, 
the space ${\cal M}$ represents the coset space 
$O(L,L-N)$, where $O(L,L-N)$ is the  
irreducible symmetric Riemannian space  
$SO(L)/S(O(N) \times O(L-N))$.\cite{helgason}
We remind here that $O(N)$ denotes the group of generic orthogonal matrices,
whereas $SO(N)$ represents the group of 
orthogonal matrices with determinant one.
Similarly $O(N) \times O(L-N)$ represents the group of block diagonal 
matrices with $N\times N$ and $L-N \times L-N$ blocks, where each block is 
in turn an orthogonal matrix.  Also the symbol $S(O (N) \times O(L-N))$
indicates that the determinant of this block diagonal matrix (the products of the determinant of each block, 
equal to $\pm1$ as for any orthogonal matrix) has to be $1$. 

This space ${\cal M}$ is compact (all the $ N (L-N)$ independent variables  represent
essentially angles of unit vectors in $L$ dimensional space) 
and there exist a unique (up to a constant) 
measure $d \mu$
 such that $d  \bar U \mu = d \mu$ for all $\bar U \in SO(L)$ 
where $SO(L)$ is the group of $L \times L$ orthogonal  matrices with unit determinant\cite{helgason}, namely $| \bar U | =1$. An  orthogonal  matrix $U$,   
acts on $|\psi \rangle $ in an obvious way, 
namely $ \psi\to \psi U$ in Eq.(\ref{slater}).
The space ${\cal M}$ can be therefore  represented by an irreducible 
symmetric Riemannian 
space. Using a matrix $U \in SO(L)$ 
we have essentially 
$L$ orthonormal directions (e.g. the raws of the matrix), and the first $N$ spans all possible Slater determinants in the space $ {\cal M}$. 
For the previous discussion 
this Slater determinant will be left unchanged (up to a sign)  
if we multiply the 
matrix $U$ for an arbitrary element of the $ S(O(N)\times O(L-N))$ 
unitary group, and therefore $ {\cal M}$ is equivalent to the space  
 $SO(L) / S(O(N) \times O(L-N))$. 

As  a further proof that ${\cal M}$ is equivalent to 
 $SO(L)/S(O(N)\times O(L-N))$ 
it is also easy to verify that 
the dimension of this space space is exactly $N (L-N)$.
The dimension of an orthogonal  matrix of dimension $D$ is $D (D-1)/2$,  
and therefore the dimension of
the coset space  $SO(L) /S(O(N)\times O(L-N))$ is $L(L-1)/2 - (L-N)(L-N-1)/2 
-N(N-1)/2= N (L-N) \blacksquare.$   

In order to represent the irreducible space $SO(L)/S(O(N)\times O(L-N))$ 
for $L >> N$,  with $N (L-N)$ 
variables, 
a possible choice is to define an unconstrained  $N \times (L-N)$ 
matrix $V$ and the corresponding unitary $L\times L$ matrix $Q$:
\begin{equation}
Q = \left( 
\begin{array}{cc}
\sqrt{ I - V V^\dag} & V \\
-V^\dag & \sqrt{ I - V^\dag V} 
\end{array}
\right)
\end{equation}
with the constraint that the positive definite matrix $V V^\dag$ has all 
eigenvalues bounded by one, namely $ V V^\dag \le 1$.
Thus we explicitly see that the space is compact.
As mentioned before we can identify a wavefunction $\psi \in M$ with the 
first $N$ raws of this unitary matrix $Q$, up to a sign, 
so that the orbitals of the 
determinant are:
\begin{equation} \label{chart}
\psi_{l,k} = Q_{l,k} ~{\rm for }~l=1,2,\cdots,N.
\end{equation}
A measure $d\psi$ of the coset (reducible) Riemann space $SO(L) / S(O(N) \times O(L-N)) $ is said to be an invariant measure 
when it remains invariant under all unitary transformations $ U \in U(L)$,
namely $ d \psi U = d \psi$.  
An invariant measure 
represented by the matrix $V$ is given by:
\begin{equation}
d \psi = C d \mu(V)
\end{equation}
where $C$ is an appropriate normalization constant, 
and $\mu(V)$ is the invariant measure in 
$SU(L) / S(U(N) \times U(L-N))$.\cite{helgason}
Although explicit formulas are known for the invariant measure, 
they look a bit complicated to be implemented in practice.
We are confident that a very convenient expression of the invariant measure is 
possible in terms of the eigenvalues of $V V^\dag$, which should amount 
to only $N^3$ operations. This would lead immediately to a computationally 
affordable extension of our method to DFT or mean-field type of ansatz.





%

\end{document}